\documentclass[%
 aip,
 amsmath,amssymb,
 reprint,%
]{revtex4-1}

\usepackage{graphicx}
\usepackage{dcolumn}
\usepackage{bm}

\usepackage[utf8]{inputenc}
\usepackage[T1]{fontenc}
\usepackage{mathptmx}

\begin{document}


\title{Tuning energy barriers by doping 2D group-IV monochalcogenides}

\author{Albert Du}
\author{Zachary Pendergrast}
\author{Salvador Barraza-Lopez}
 \email{sbarraza@uark.edu}
\affiliation{Department of Physics, University of Arkansas, Fayetteville, AR 72701, USA}

\date{\today}

\begin{abstract}
Structural degeneracies underpin the ferroic behavior of next-generation two-dimensional materials, and lead to peculiar two-dimensional structural transformations under external fields, charge doping and/or temperature. The most direct indicator of the ease of these transformations is an {\em elastic energy barrier}, defined as the energy difference between the (degenerate) structural ground state unit cell, and a unit cell with an increased structural symmetry. Proximity of a two-dimensional material to a bulk substrate can affect the magnitude of the critical fields and/or temperature at which these transformations occur, with the first effect being a relative charge transfer, which could trigger a structural quantum phase transition. With this physical picture in mind, we report the effect of modest charge doping (within $-0.2$ and $+0.2$ electrons per unit cell) on the elastic energy barrier of ferroelastic black phosphorene and nine ferroelectric monochalcogenide monolayers. The elastic energy barrier $J_s$ is the energy needed to create a $Pnm2_1\to P4/nmm$ two-dimensional structural transformation. Similar to the effect on the elastic energy barrier of ferroelastic SnO monolayers, group-IV monochalcogenide monolayers show a tunable elastic energy barrier for similar amounts of doping: a decrease (increase) of $J_s$ can be engineered under a modest hole (electron) doping of no more than one tenth of an electron or a hole per atom.
\end{abstract}

\maketitle

\section{Introduction}\label{sec:intro}
In the very recent past, Zhu, Lu, and Wang showed that the intrinsic electric dipole of SnSe monolayers can be tuned by charge doping.\cite{Zhu} Charge doping is a variant of charge density reaccommodation, with another instance of electronic rearrangement, dipole screening, and structural modifications being created by illumination from light.\cite{haleoot_prl_2017_sns_snse} Reference \onlinecite{Zhu} employed the Perdew-Burke-Ernzerhof (PBE) approximation for exchange and correlation\cite{PBE} in density functional theory.\cite{martin}

According to Seixas {\em et al.} \cite{Seixas2016} and others,\cite{arxiv,adp} charge doping also modifies energy barriers separating the ground state ferroelectric unit cell and paraelectric unit cells that have an enhanced symmetry. As indicated by Potts,\cite{potts} a change in energy barriers in turn modifies the critical temperature at which structural phase transformations take place in ferroic 2D materials. Thus far, only chemical composition,\cite{Mehboudi2016,shiva} strain,\cite{other4} and structural constraints\cite{newarXiv} have been studied as means to control energy barriers of group-IV monochalcogenides, and this manuscript shows that energy barriers are also susceptible of change upon charge doping, which could occur by proximity to a supporting substrate.\cite{Kai,KaiPRL,sntebl,bulk2}

In doing so, we recall that density functional theory methods are unable to describe the electron correlation in so-called ``van der Waals solids''  accurately, as explicitly shown in the bulk and in bilayers of black phosphorus.\cite{shulenburger} One may conclude that this may also be the case for isoelectronic group-IV monochalcogenide monolayers. Previous work from us\cite{shiva} questions the accuracy of exchange-correlation approximations such as LDA\cite{LDA,LDA2} or even PBE,\cite{PBE} and suggests that these materials could become a testbed for further work in exchange-correlation functionals. In that previous study, geometries were determined from density-functional theory \cite{martin} using eight different exchange-correlation (XC) functionals that include traditional ones (LDA \cite{LDA,LDA2} and PBE \cite{PBE}), five with self-consistent van der Waals corrections \cite{Becke,dion2004,reviewvdw} (optPBE-vdW \cite{klimes1,klimes2}, optB86b-vdW \cite{klimes1,klimes2}, vdW-DF-cx \cite{cx}, vdW-DF2 \cite{DF2}, B86R-vdW-DF2 \cite{DF2,Hamada}) and the recently developed SCAN+rVV10 \cite{scan}, which has been successful to describe the weak bonding in liquid and solid water in the most precise manner yet \cite{PNASscan}. In order to test the predictions obtained with the PBE exchange-correlation functional, here we worked with the B86R-vdW-DF2\cite{DF2,Hamada} and with a combination of optPBE exchange and DF2 correlation corrections (optPBE-vdW-DF2); these choices were made for purely illustrative purposes only.

The atomistic structure of black phosphorus monolayers and most group-IV monochalcogenide monolayers is peculiar in that a finite horizontal tilt $\delta_{x,0}$ (or $d$ in Ref.~\onlinecite{Zhu}) exists in between pairs of atoms (exceptions are PbS, PbSe, and PbTe for which $\delta_{x,0}=0$\cite{Mehboudi2016,shiva}). Additional variables that permit understanding the structural evolution with charge doping are lattice parameters $a_{1,0}$ and $a_{2,0}$\cite{Mehboudi2016,other4,shiva,newarXiv} (labeled $a$ and $b$ in Ref.~\onlinecite{Zhu}). The angle $\alpha$ in Ref.~\onlinecite{Zhu} is not a good descriptor of a paraelectric structure.

Finally, the process to obtain energy barriers under doping is straightforward, but additional steps beyond those described in Ref.~\onlinecite{Zhu} are necessary. The main difference is that while the focus of Ref.~\onlinecite{Zhu} is on the ground state unit cell which has a Pnm$2_1$ symmetry,\cite{rodin_prb_2016_sns} the barrier $J_s$ to be calculated as energy differences among such ground state unit cell and a paralectric unit cell with P4/nmm symmetry.

The manuscript is straightforward, it has a decisive emphasis on atomistic structure, and it is organized as follows: Computational details are provided in Sec.~\ref{sec:ii}, a comparative discussion that includes the results from Zhu and coworkers obtained with the PBE approximation to exchange correlation and ours, and the calculation of energy barriers is provided in Sec.~\ref{sec:iii}. Conclusions are provided in Sec.~\ref{sec:iv}.

\begin{figure*}[tb]
\begin{center}
\includegraphics[width=0.96\textwidth]{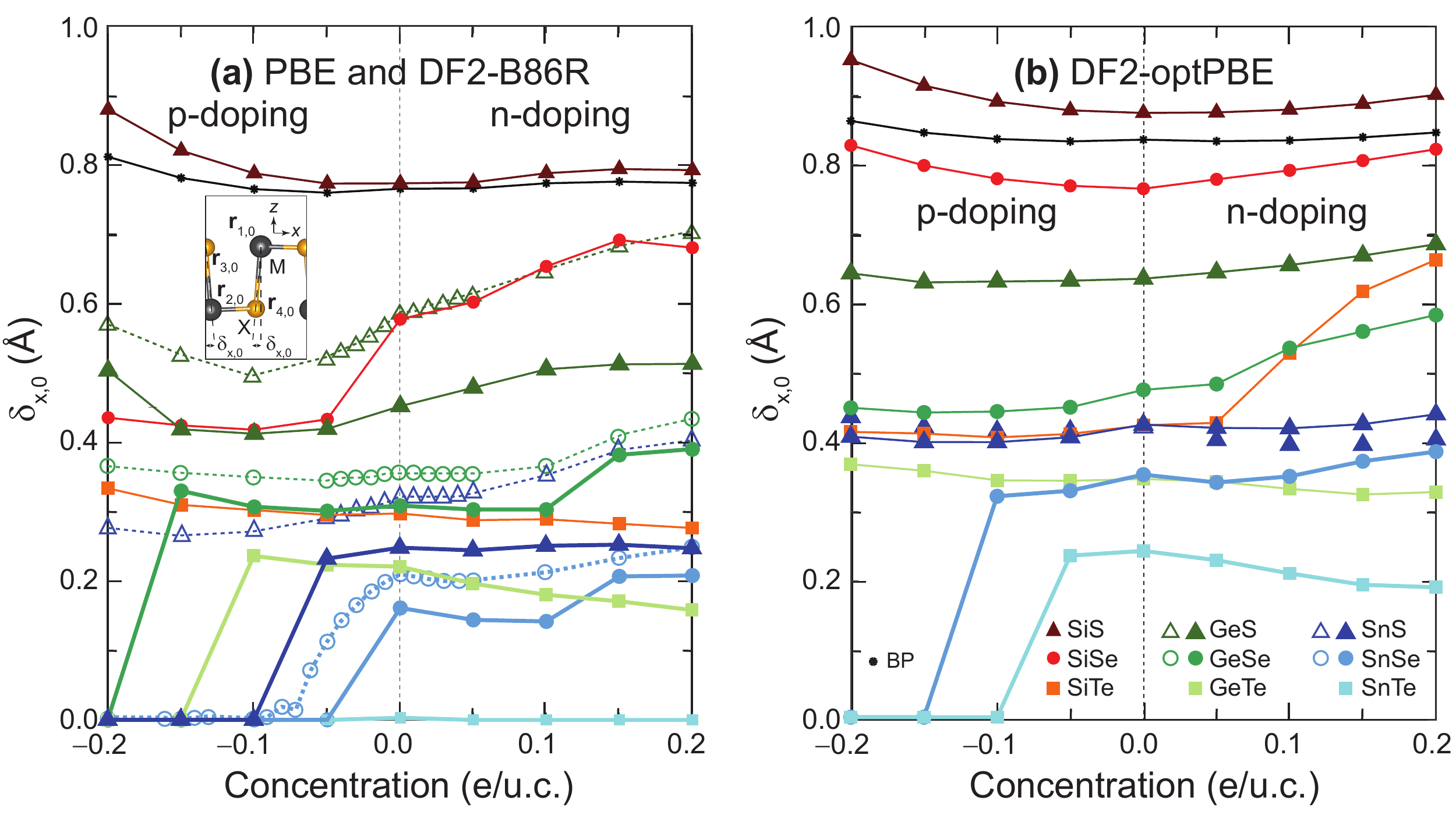}
\caption{Order parameter $\delta_{1,0}$ as a function of hole/electron concentration for black phosphorus monolayer and nine group-IV monochalcogenide monolayers (SiS, SiSe, SiTe, GeS, GeSe, GeTe, SnS, SnSe and SnTe). (a) Results with the DF2-B86R exchange-correlation functional are shown in solid symbols, and PBE results from Ref.~\onlinecite{Zhu} can be seen as open symbols. GeSe, GeTe, SnS, SnSe, and SnTe monolayers are paraelectric with a modest hole doping of 0.2 holes/u.~c. (b) As the alternate calculation employing the DF2-optPBE functional shows, the magnitude of $\delta_{x,0}$ is strongly dependent on exchange-correlation functional: Here, SnSe and SnTe monolayers become paraelectric under a doping of 0.2 holes/u.c.}\label{fig:fig1}
\end{center}
\end{figure*}

\section{Computational details}\label{sec:ii}

Calculations were performed with the VASP code \cite{vasp} (release 5.4.4) on a 30$\times$30$\times$1 $k-$point mesh and with a 600 eV energy cutoff. Energy and force convergence criteria were set to $10^{-11}$ eV and $10^{-5}$ eV/\AA{} respectively, and the high precision tag was turned on. The out-of-plane lattice vector length was 30 \AA{}. The anharmonicity of the energy landscape of monolayers makes it difficult for standard algorithms that optimize lattice vectors to find the overall minima. We have therefore performed calculations on preestablished lattice parameter meshes (i.e., in meshes for which the variation of energy against lattice parameters is sampled with a 0.005~\AA{} resolution and the four basis atoms are allowed to move along the $x-$ and $z-$directions).

\section{Results and discussion}\label{sec:iii}

Figure \ref{fig:fig1} shows the evolution of zero-temperature $\delta_{x,0}$ {\em versus} charge doping; the inset of Fig.~\ref{fig:fig1}(a) showing its schematic depiction. In the inset, the four atoms forming the unit cell are shown; the gray atom is the group-IV (metal, M) element and the yellow one the chalcogen (X).

\begin{figure*}[tb]
\begin{center}
\includegraphics[width=0.96\textwidth]{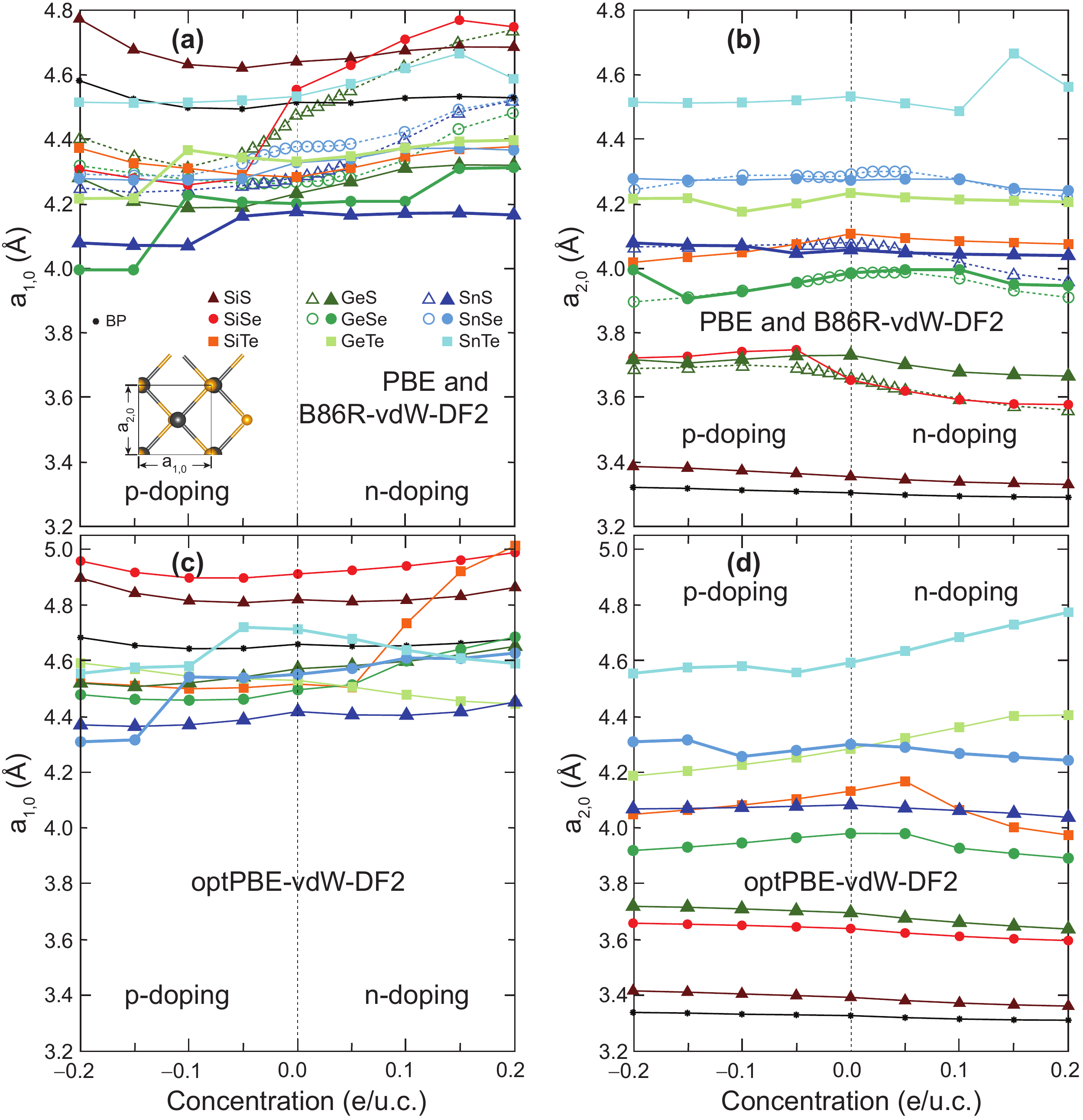}
\caption{Lattice parameters $a_{1,0}$ (subplots (a) and (c)) and $a_{2,0}$ (subplots (c) and (d)) as a function of charge doping concentration. Results in subplots (a) and (b) were obtained with the DF2-B86R exchange-correlation functional, and open symbols correspond to calculations with the PBE exchange correlation functional in Ref.~\onlinecite{Zhu}.  Note that heavier monochalcogenide monolayers (GeSe, GeTe, SnS, SnSe, and SnTe) have identical lattice parameters for the largest shown hold doping. Subplots (c) and (d) show data obtained with the DF2-optPBE exchange-correlation functional. In this case, only SnSe and SnTe monolayer show converging lattice parameters for the largest hole doping shown in these subplots.}\label{fig:fig2}
\end{center}
\end{figure*}

The B86R-vdW-DF2 functional in Fig.~\ref{fig:fig1}(a) performs in a manner analog to PBE in the four materials (GeS, GeSe, SnS and SnSe) studied previously\cite{Zhu} (shown in dashed lines and open symbols). We can attest to the accuracy of the charge-neutral structures in Ref.~\onlinecite{Zhu}, as our PRB results are spot-on\cite{shiva} when compared with theirs.

\begin{figure*}[tb]
\begin{center}
\includegraphics[width=0.96\textwidth]{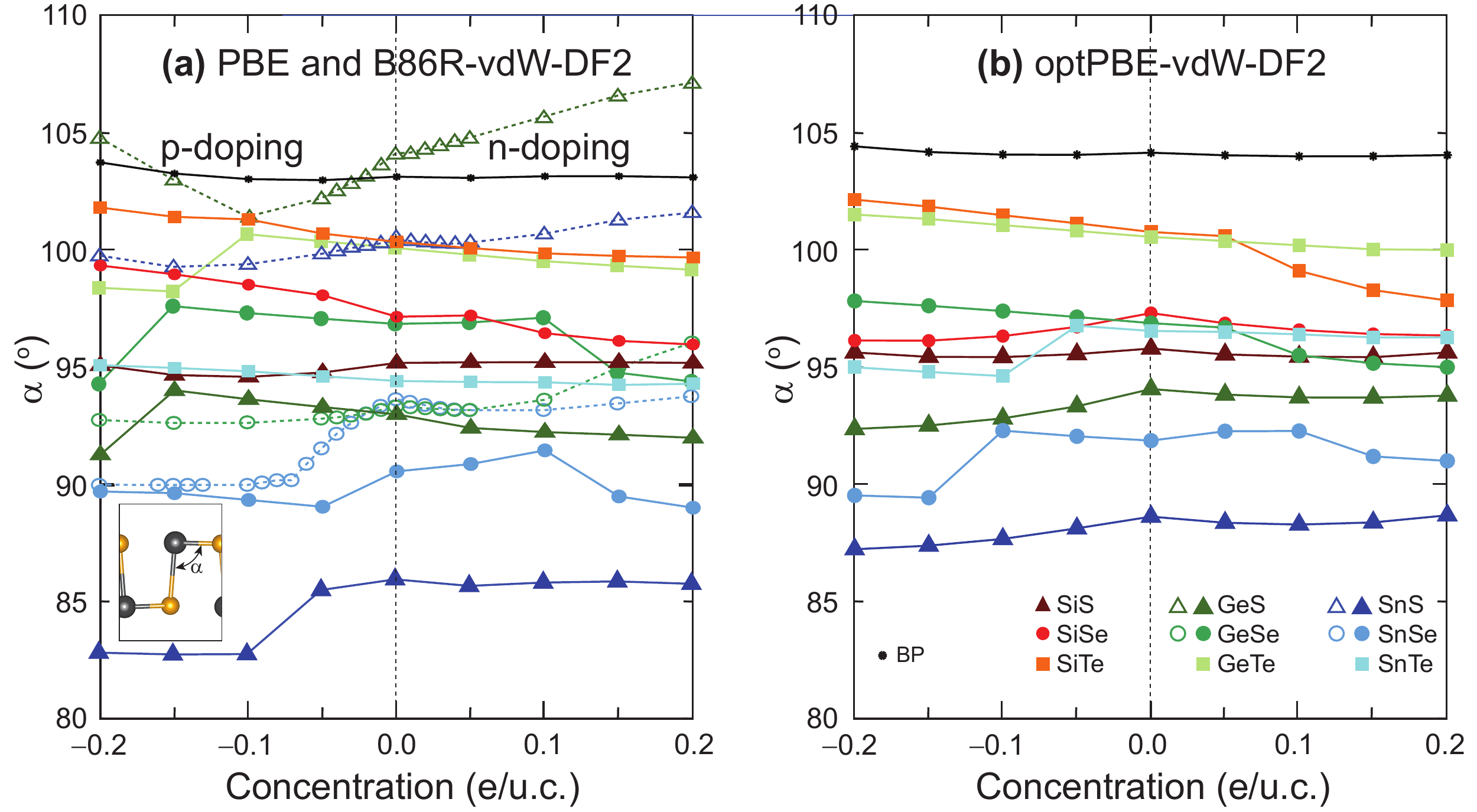}
\caption{The angle $\alpha$ is not a reliable order parameter of a paraelectric u.c.~as a function of charge doping: its magnitude does not necessarily correlate with $\delta_{x,0}=0$ in Fig.~\ref{fig:fig1}. Furthermore, $\alpha$ can take smaller or larger values than 90$^{\circ}$. (a) Results with the DF2-B86R exchange-correlation functional are shown in solid symbols, and PBE results from Ref.~\onlinecite{Zhu} can be seen as open symbols. (b) Alternate calculation employing the DF2-optPBE exchange-correlation functional.}\label{fig:fig3}
\end{center}
\end{figure*}

The B86R-vdW-DF2 exchange-correlation functional provides a structure slightly compressed with respect to PBE, as will be seen more clearly on Fig.~\ref{fig:fig2} when lattice parameters are revealed. Results with the B86R-vdW-DF2 functional provide slightly {\em smaller} magnitudes of $\delta_{x,0}$ when compared to previous PBE results. Additionally, there is a salient difference in our results and previous ones: while we confirm that SnSe monolayers are paraelectric under hole doping, {\em we also observe SnTe, SnS, GeTe, and even GeSe to turn paraelectric under hole doping} (previous work does not report GeSe, GeTe, nor SnS to be paraelectric). Occurring at zero temperature, {\em the change from a ferroelectric to a paraelectric ground state structure upon doping is a quantum phase transition}.\cite{sachdev}

As seen in Fig.~\ref{fig:fig1}(b), the magnitude of $\delta_{x,0}$ obtained when employing the optPBE-vdW-DF2 functional is larger than that observed with PBE or B86R-vdW-DF2. This is due to the variation in structure observed among many exchange-correlation functionals as reported in Ref.~\onlinecite{shiva}. The larger magnitude of $\delta_{x,0}$ is such that only SnSe and SnTe monolayers can become paraelectric under the amount of charge doping shown in the Figure.

We now turn our attention to the evolution of zero-temperature lattice parameters $a_{1,0}$ and $a_{2,0}$ as a function of charge doping for black phosphorus and the nine group-IV monochalcogenide monolayers. For this purpose, the left subplots in Fig.~\ref{fig:fig2} display $a_{1,0}$, while subplots to the right show $a_{2,0}$. Once again, it is possible to see the close correspondence among B86R-vdW-DF2 and PBE results, and the larger values (especially of $a_{1,0}$ when the optPBE-vdW-DF2 exchange correlation functional is employed. We chose to give an identical range for both lattice parameters in order to emphasize that they can take on identical values for those cases in which $\delta_{x,0}$ turned zero in Fig.~\ref{fig:fig1}.

\begin{figure*}[tb]
\begin{center}
\includegraphics[width=0.96\textwidth]{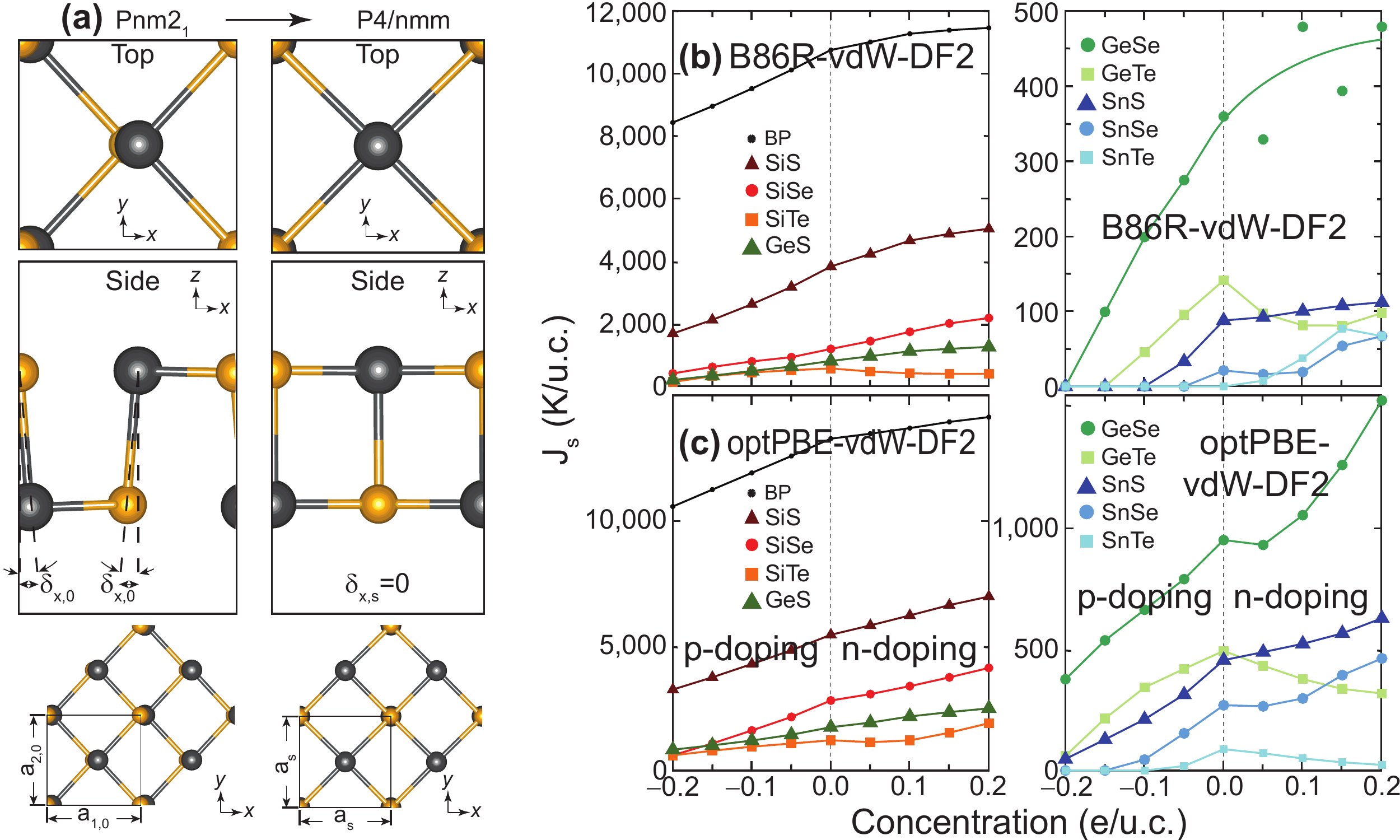}
\caption{(a) Geometrical depiction of the Pnm2$_1 \to$ P4/nmm structural transformation, leading to the barrier $J_s$. (b) Energy barriers $J_s$ as obtained for the DF2-B86R exchange correlation functional. (c) Energy barriers $J_s$ as obtained for the DF2-optPBE exchange correlation functional. There is a clear decrease (increase) of $J_s$ with hole (electron) doping.}\label{fig:fig4}
\end{center}
\end{figure*}

The angle $\alpha$ in Ref.~\onlinecite{Zhu} is formed among atoms $\mathbf{r}_{3,0}$, $\mathbf{r}_{2,0}$, and $\mathbf{r}_{4,0}$ at the inset of Fig.~\ref{fig:fig1}(a). Using their atomic positions as defined before,\cite{other4,shiva}
\begin{eqnarray*}
\mathbf{r}_{3,0}=(0,0,z_{3,0})\text{, }\mathbf{r}_{2,0}=(\delta_{x,0},0,0)\text{, and}\\
\mathbf{r}_{4,0}=\left(\frac{a_{1,0}}{2},\frac{a_{2,0}}{2},z_{z,0}-z_{3,0}\right),
\end{eqnarray*}
with $z_{1,0}$ and $z_{3,0}$ relative heights with respect to atom $\mathbf{r}_{2,0}$ (which is placed at a zero height), it becomes possible to write down a compact expression, as follows:
\begin{equation}\label{eq:eq1}
\cos\alpha=\frac{-\delta_{x,0}\left(\frac{a_{1,0}}{2}-\delta_{x,0}\right)
+z_3(z_1-z_3)}{\sqrt{\delta_{x,0}^2+z_{3,0}^2}
\sqrt{\left(\frac{a_{1,0}}{2}-\delta_{x,0}\right)^2+
\left(\frac{a_{2,0}}{2}\right)^2+(z_{1,0}-z_{3,0})^2}}.
\end{equation}

Equation \eqref{eq:eq1} gives important information away. The first term in its numerator is smaller or equal than zero, as $0\leq\delta_{x,0}< a_{1,0}$. In turn, the second term in its numerator is larger or equal than zero. In particular, and as indicated as early as 2016, the relative height of the lowermost atoms $z_{1,0}-z_{3,0}$ {\em can be negative or positive depending on chemical compound,}\cite{kamal_prb_2016_iv_vi_monolayers} while it appears that Ref.~\onlinecite{Zhu} assumes it to be zero (as that would be the only way $\alpha\geq 90^{\circ}$ in that work). In other words, and as seen in Fig.~\ref{fig:fig3}, $\alpha$ should not be employed as the order parameter to signal the ferroelectric to paraelectric quantum phase transformation.

It is time to shift gears and show how the elastic energy barrier $J_s$ is influenced by charge doping.  For this purpose, Fig.~\ref{fig:fig4}(a) displays the ferroelectric orthorhombic ground state unit cell with Pnm2$_1$ group symmetry, and the paraelectric tetrahedral unit cell with P4/nmm group symmetry from which $J_s$ is computed.\cite{Mehboudi2016,shiva} The units of $J_s$ (K/u.c.) are directly proportional to the critical temperature $T_c$ at which the structural transformation takes place.\cite{newarXiv} Figures \ref{fig:fig4}(b) and \ref{fig:fig4}(c) show that the barrier is smaller with hole doping, and it gradually increases to take largest values with electron doping for a certain degree of tunability. Obeying to the largest magnitudes of both $\delta_{x,0}$ and $a_{1,0}$ reported in Figs.~\ref{fig:fig1} and \ref{fig:fig2}, the barrier is larger when employing the optPBE-vdW-DF2 functional. As indicated before, $J_s$ is the first estimate of a possible critical temperature at which the ferroelectric to paraelectric structural transformation might take place,\cite{Mehboudi2016,newarXiv} and hence the significance of the tunability observed in Fig.~\ref{fig:fig4}. The fact that the barriers increase from hole to electron doping give confidence that such phenomena may be independent of the exchange-correlation functional being employed and is thus a reliable feature of these materials.

\section{conclusions}\label{sec:iv}
To conclude, we studied the effect of charge doping (within $-0.2$ and $+0.2$ electrons per unit cell) on the elastic energy barrier created by a Pnm2$_1 \to$ P4/nmm two-dimensional structural transformation of ferroelastic black phosphorene and nine ferroelectric monochalcogenide monolayers, using DF2-B86R and DF2-optPBE exchange-correlation functionals for this purpose. . The previous study\cite{Zhu} does not report energy barriers $J_s$. These barriers are crucial to determine the structural transformation. We studied many more materials than they did, and the effect of charge doping on these barriers we had never assessed before.

Providing a comparison against recent results published in this Journal, the zero-temperature evolution of the in-plane tilt $\delta_{x,0}$, an angle $\alpha$, and lattice parameters $a_{1,0}$ and $a_{2,0}$ of the ground state unit cell were provided along the way.

Group-IV monochalcogenide monolayers show a tunable elastic energy barrier for similar amounts of doping: a decrease (increase) of $J_s$ can be engineered under a modest hole (electron) doping of no more than one tenth of an electron or a hole per atom, in addition to their tunability by the choice of chemical formula and of exchange-correlation functional being employed. These results--{\em strictly on the atomistic structure}--provide further guidance concerning a possible tunability of the critical temperature of these compounds by charge doping.

The data that support the findings of this study are available from the corresponding author upon reasonable request.

\begin{acknowledgments}
A.D. was funded by the NSF through a Research Experience for Undergraduates (REU) program at the University of Arkansas, Grant DMR-1851919. Z.P. received support from NSF, Grant DMR-1610126. S.B.L. was funded by an Early Career Grant from the DOE (Award  DE-SC0016139). Calculations were performed on Cori at NERSC, a U.S. DOE Office of Science User Facility operated under Contract No. DE-AC02-05CH11231, and on Trestles and Pinnacle at Arkansas, funded by the U.S. National Science Foundation (Grants 0722625, 0959124, 0963249, and 0918970), a grant from the Arkansas Economic Development Commission, and the Office of the Vice Provost for Research and Innovation.
\end{acknowledgments}


%

\end{document}